\def\year{2022}\relax
\documentclass[letterpaper]{article} 
\usepackage{aaai22}  
\usepackage{times}  
\usepackage{helvet}  
\usepackage{courier}  
\usepackage[hyphens]{url}  
\usepackage{graphicx} 
\usepackage{natbib}  
\usepackage{caption} 
\DeclareCaptionStyle{ruled}{labelfont=normalfont,labelsep=colon,strut=off} 
\usepackage{algorithm}
\usepackage{algorithmic}
\usepackage{amsmath,color}
\usepackage{mathtools}
\usepackage{newfloat}
\usepackage{listings}
\floatstyle{ruled}
\newfloat{listing}{tb}{lst}{}
\floatname{listing}{Listing}
\usepackage{color} 

\title{Cyberattack Detection in Critical Infrastructure and Supply Chains}
\author{
    Smita Khapre
}
\affiliations{
    Department of Engineering and Applied Science\\

    1420 Austin Bluffs Pkwy\\
    Colorado Springs, Colorado 80918\\
}

\usepackage{bibentry}

\begin{document}

\maketitle

\begin{abstract}
Cyberattack detection in Critical Infrastructure and Supply Chains has become challenging in Industry 4.0. Intrusion Detection Systems (IDS) are deployed to counter the cyberattacks. However, an IDS effectively detects attacks based on the known signatures and patterns, Zero-day attacks go undetected. To overcome this drawback in IDS, the integration of a Dense Neural Network (DNN) with Data Augmentation is proposed. It makes IDS intelligent and enables it to self-learn with high accuracy when a novel attack is encountered. The network flow captures datasets are highly imbalanced same as the real network itself. The Data Augmentation plays a crucial role in balancing the data. The balancing of data is challenging as the minority class is as low as 0.000004\% of the dataset, and the abundant class is higher than 80\% of the dataset. Synthetic Minority Oversampling Technique is used for balancing the data. However, higher accuracies are achieved with balanced test data, lower accuracies are noticeable with the original imbalanced test data suggesting overfitting. A comparison with state-of-the-art research using Synthetic Minority Oversampling Technique with Edited Nearest Neighbor shows the classification of classes remains poor for the original dataset. This suggests highly imbalanced datasets of network flow require a different method of data augmentation.
\end{abstract}

\section{Introduction}
In Industry 4.0, Cyber-physical systems (CPS) \cite{sha2015data, malik2023developing}, are everywhere in critical infrastructure. They communicate using the wireless network or short-distance communication technology. IDS, before Industry 4.0, did reasonably well in detecting cyberattacks. With 5G, the CPS has become the norm in the supply chain and critical infrastructure. The CPS applications in these sectors, introduced security issues and challenges \cite{ravi2022recurrent}. IDS can no longer cope with the cyber-threats unless it becomes intelligent and learns from past data. It needs to identify the new patterns as threats. Much research has been carried out on how to make IDS intelligent for the Internet of Things (IoT), Industrial IoT, Cloud computing, and Supply Chain. This work proposes to consolidate the different approaches and evaluate them. Deep Learning based approaches in \cite{abd2023intrusion, awajan2023novel, chauhdary2023efficient, jayesh2023hybrid, kummerow2023cyber, pujol2022unveiling, ravi2022recurrent, gupta2021lio}, target their integration with IDS, to enhance the ability to detect Zero-day attack and mitigate them. At the same time, newer research in Neural Networks will potentially add to the accuracy and performance. Nevertheless, the transformer architecture is believed to add better value as it can extract temporal features with the multihead attention mechanism. 
While the transformer Neural Networks research and development is not in the scope of the proposed work, it lays the foundation. 

CICIDS-2017 Dataset in \cite{sharafaldin2018CICIDS-2017} is generated in testbed architecture in 5 days. The generated packet data is captured and converted to network flows. And their relevant features are extracted using CICFlowMeter. It consists of 7  major attack categories. They include Benign, DDoS, DoS, PortScan, Patator, Web Attacks, Bot, and Infiltration classes network flows.  It is open-source and available publicly for research to train and test the model design. The models developed using this dataset help in circumventing various challenges like missing data, data imbalance, feature extraction, and classification.

The section \textit{Background} discusses the motivation of this project. The section \textit{Related Work} contains a summary of related research. \textit{Dataset} gives a brief history of CICIDS-2017 dataset and its relevance. The section \textit{Evaluation} mentions the dataset details and evaluation metrics. The section \textit{Approach} elaborates on the proposed design and experiments. The section \textit{Results} elaborates on the experiment results and comparison. Finally, the sections \textit{Conclusion} and \textit{Future Work} conclude the project. 

\section{Background}
With advancements in wireless technologies, CPS has taken over the critical infrastructure industry. It paved the way for cyberattacks. The cybercriminal attacks them to gain control over them with various malicious intents. Any solution deployed to counter them goes obsolete much sooner than new solutions are implemented. There is a dire need to replace them with self-learning systems that can address zero-day attacks. Artificial Neural Network (ANN) has the potential to play a vital role. It is understood that there is no perfect solution to cyberattacks in evolving technology. The latest technological skills are used in defense and attacks simultaneously. New hardware, software, and attacks are developed simultaneously, making the defense lag behind. Thus, there is a need to integrate ANNs in the IDS, which updates itself based on a new set of parameters \cite{fausett2006fundamentals}. This will make the cyber-defense dynamic, persistent, and resilient.

The implementation of ANNs in various applications had exponential growth recently. It is prevalent in home automation, automotive, industrial automation, flight simulation, healthcare, traffic control systems, transportation, aviation, and defense are the least to mention. With the advent of Transformer Architecture in 2017, by \cite{vaswani2017attention}, there has been a revolution in their research and applications. There are multiple papers on IDS based on the transformer capable of extracting temporal features without the complexity of Recurrent Neural Networks (RNN). Transformer networks are also time-efficient compared to RNNs. 

\section{Related Work}

Intrusion Detection System using transformer-based
transfer learning for Imbalanced Network Traffic (IDS-INT) proposed by \cite{ullah2023ids} to identify a specific attack having complex features and data imbalance issues. They used the Synthetic Minority Oversampling Technique (SMOTE) to balance abnormal traffic
and detect minority attacks. The order of activities in this work is categorization, transformer-transfer-learning, SMOTE, CNN then prediction on the processed dataset. The overall accuracies and F1-scores achieved for the proposed model lie between 99\% to 100\%. The datasets used are UNSW-NB15, NSL-KDD, and CICIDS-2017.

 "Transformer and Bidirectional Long Short-Term Memory" is proposed by \cite{gan2023research} for intrusion detection. Deep Neural Network layers and a softmax layer were deployed for extracting features and classification. The order of activities in this work is Data preprocessing, train-and-test split, training on the transformer(encoder-only), LSTMs, and DNN model, and then testing on the preprocessed dataset. The overall accuracies achieved for different classes lie between 90\% to 99\%. The dataset used is NSL-KDD.

"An intrusion detection model
using ResNet, Transformer, and BiLSTM (Res-TranBiLSTM)" is proposed by \cite{wang2023res-TransBiLSTM_imb}, utilizing both the spatial and temporal features of network traffic. They have used Synthetic Minor Overriding Technique Edited Nearest Neighbor (SMOTE-ENN) method for solving the data imbalance problem. They argued SMOTE algorithm has the disadvantage of overlapping the other examples of neighboring classes. SMOTE-ENN identifies the overlap and deletes them. The order of activities is Digitization, Balancing, Normalization, and proposed Model training.  The datasets used are NSL-KDD and CIC-IDS2017. In this work, the testing performance on preprocessed data and original data is compared showing huge variation on CICIDS-2017. For example, the web attacks F1-score is original data is 8.64\% whereas processed data is 98.7\%. It compared the performance on different datasets and different architectures as well.

"Parallel Cross Convolutional Neural Network (PCCN)" proposed by \cite{zhang2019pccn} to deal with the multi-class Network Flows Imbalance. They argued that the use of SMOTE will increase the data and will require excessive time to train. This work uses a different feature selection algorithm which yields only 12 attack classes instead of 14, and dropped the Benign class from the dataset altogether. Interestingly, their algorithm yielded a higher number of examples than the original dataset.  

"RTIDS: A Robust Transformer-Based Approach for Intrusion Detection System" by \cite{wu2022rtids}, the order of activities is data cleaning, data normalization, and feature selection 
and dataset splitting before model training. They have used SMOTE to increase the volume of minority classes. the F1-Scores and accuracies for all classes reached 99\% except for the minority classes namely Heartbleed, SQL injection, and Bot, suggesting the balancing issue. 
The datasets used are CICIDS2017 and CIC-DDoS2019. 

"A CNN-transformer hybrid approach for an intrusion
detection system in advanced metering infrastructure" by \cite{yao2023cnn}, used the Adaptive Synthetic sampling (ADASYN) technique for data balancing. Their design model is based on CNN and Transformer. The order of activities is feature ranking using XGBoost, feature screaming, data balancing, digitization and one-hot encoding, normalization then the training model. 
The overall accuracies reported on their design are 91.04\%, 97.85\%, and 91.06\% on the datasets NSL-KDD, KDDCup99, and CICIDS-2017 datasets. 

"LIO-IDS: Handling class imbalance using LSTM and improved one-vs-one
technique in intrusion detection system" by \cite{gupta2021lio} used layered classification architecture. In the first layer, it classifies Benign and Attack classes, and in the second layer, it has multi-class classifier. For data balancing, they have used Borderline-SMOTE, and SVM-SMOTE. They have used NSL-KDD, CIDDS-001, and CICIDS-2017 datasets. Their model splits the dataset into training and testing, and performs binary classification. The data-balancing is part of layer 2 classification. They have achieved overall accuracies of 96\%, 99\%, and 91\% on CICIDS-2017, CIDDS-001, and NSL-KDD datasets respectively. 

\section{Dataset}
\begin{figure}[H]
    \centering
    \includegraphics{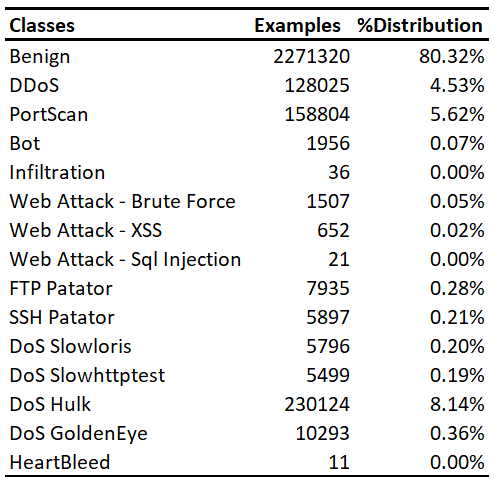}
    \caption{Class Distribution in CICIDS-2017 Dataset after missing values examples}
    \label{fig:attack-distri}
\end{figure}

CICIDS-2017 dataset is generated by \cite{sharafaldin2018CICIDS-2017}, providing the network flows dataset captured over 5 days in 8 comma-seperated-values files for various classes. All these files are consolidated into one file to form a single dataset of all classes for this project. 
CICIDS-2017 Dataset has more than 2.83 million network flows with 78 features and 15 class labels. The normal flows classified as Benign class constitute more than 80\% of the dataset. The class labels are shown in Figure \ref{fig:attack-distri}. \\

It is generated in testbed architecture with the Victim-Network and the Attack-Network, for the research community focussing on building a Machine Learning based intrusion detection system. Before 2017, the available datasets lacked the volume and similarity to real-world network flows. CICIDS-2017 provided both, and diverse attack classes too. The network flows features are extracted using CICFlowMeter, which consolidates the network packet flows between a source and destination and vice versa. It strips them of the timestamp. Instead provides inter-arrival time. Inter-arrival time is the time duration of the network packet flowing between source to destination and vice versa. Thus using this kind of dataset on simulated environments provides a good amount of learning to models. But, these models cannot be directly deployed in real-world scenarios. 

Even though it was developed in 2017 using operating systems like Windows Vista, Windows 7, Windows 8, Windows 10, Ubuntu 12, Ubuntu 14, Ubuntu 16 is now outdated, it is one of the most relevant and being used in State-of-the-arts Neural Network based IDS designs. It has flows of HTTP, SMTP, SSH, IMAP, POP3, and FTP protocols with
full packet payload. It lacks 5G protocols like HTTPS and TLS. Due to its diversity and completeness, it is widely used in IDS research to train and test neural network models and machine learning algorithms in the simulated environment. It provides real-world challenges of missing data, highly imbalanced network flows, and multi-class attacks. 

\section{Evaluation}
\begin{figure*}[htbp]
    \centering
    \includegraphics[width=16cm,height=8cm]{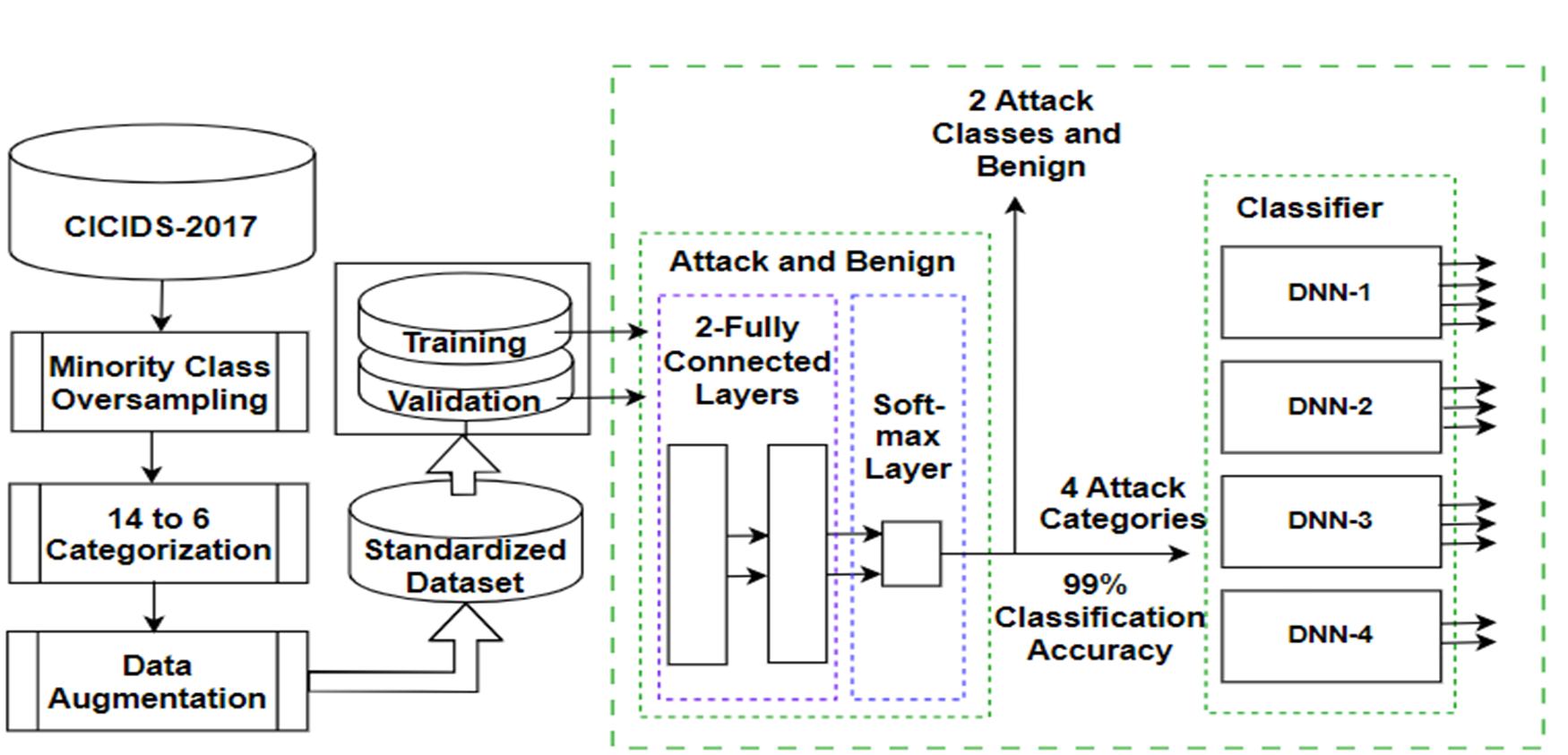}
    \caption{IDS2: Proposed Design with Categorization}
    \label{fig:IDS2}
\end{figure*}

The proposed models in Figure \ref{fig:IDS1} and \ref{fig:IDS2} are evaluated on the CICIDS-2017 Dataset for training and testing. CICIDS-2017 Dataset is an imbalanced dataset with nearly 80\% of the benign class and the 20\% flows constitute all 14 attack classes. The attack classes having a lower number of flow examples are referred to as the minority class. The attack classes close to the maximum number of flow examples are referred to as abundant classes. 

The accuracy in equation (1), is the measure of the correct prediction of class labels in total predictions. \\
The F1-Scores in equation (2), is the weighted average of the accuracy and the measure of the harmonic mean of precision and recall.
F1-Score in Figure \ref{fig:BalVsImb}, is evaluated for the classification in the designed model units. It is fine-tuned with multiple experiments to improve the F1-Score of each class.\\
In this project, F1-Scores are extensively used. It provides a measure of precision and recall indirectly.

\begin{align}
    & Accuracy = \frac{TP + TN}{TP + TN + FP + FN} \\
    & F1Score = {2 * \frac{Precision * Recall}{Precision + Recall} } \\
    & Precision = \frac{TP}{TP + FP} \\
    & Recall = \frac{TP}{(TP + FN)} 
\end{align}

To visualize true positives (TP), true negatives (TN), False Positives (FP), and False Negatives (FN), heat maps of the seaborn library are used. 

\begin{figure}[H]
    \centering
    \includegraphics[width=8cm,height=7cm]{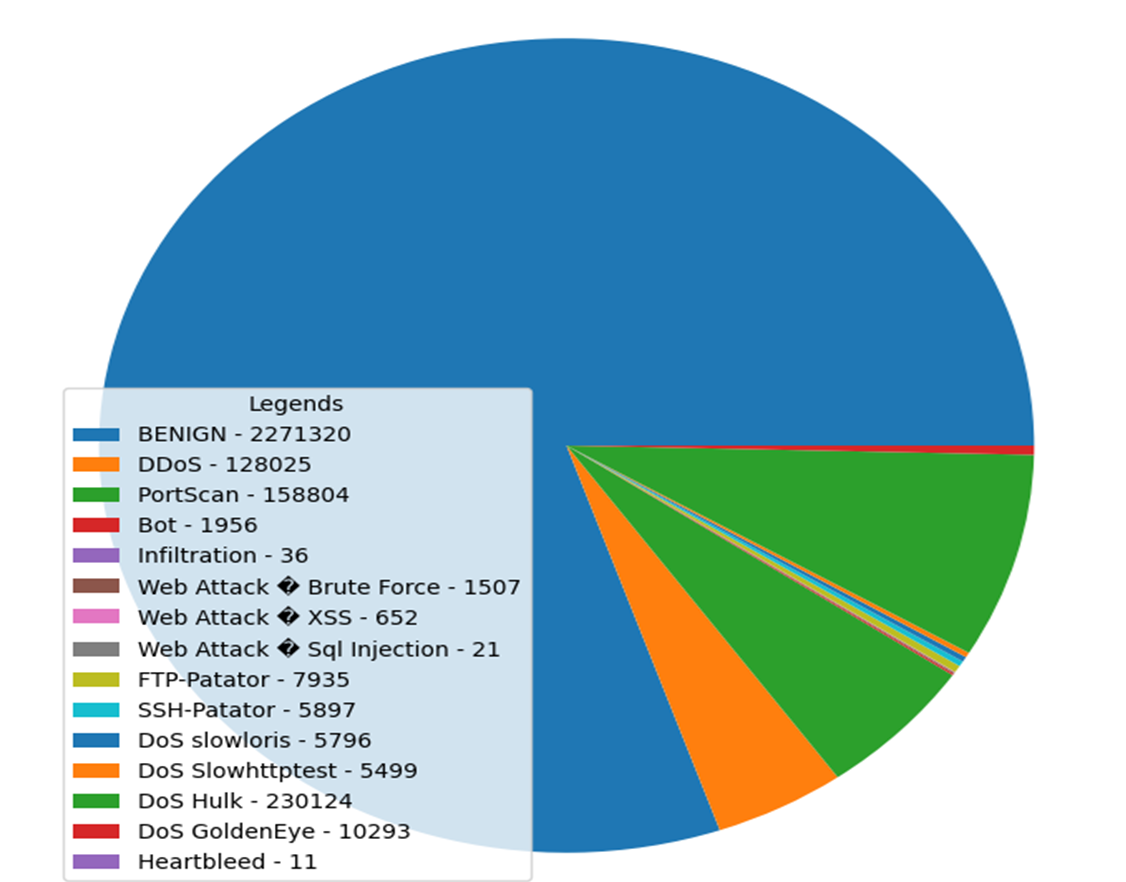}
    \caption{CICIDS-2017 Before Balancing}
    \label{fig:initialDS}
\end{figure}

\section{Approach}
\begin{figure}[H]
    \centering
    \includegraphics[width=8cm,height=7cm]{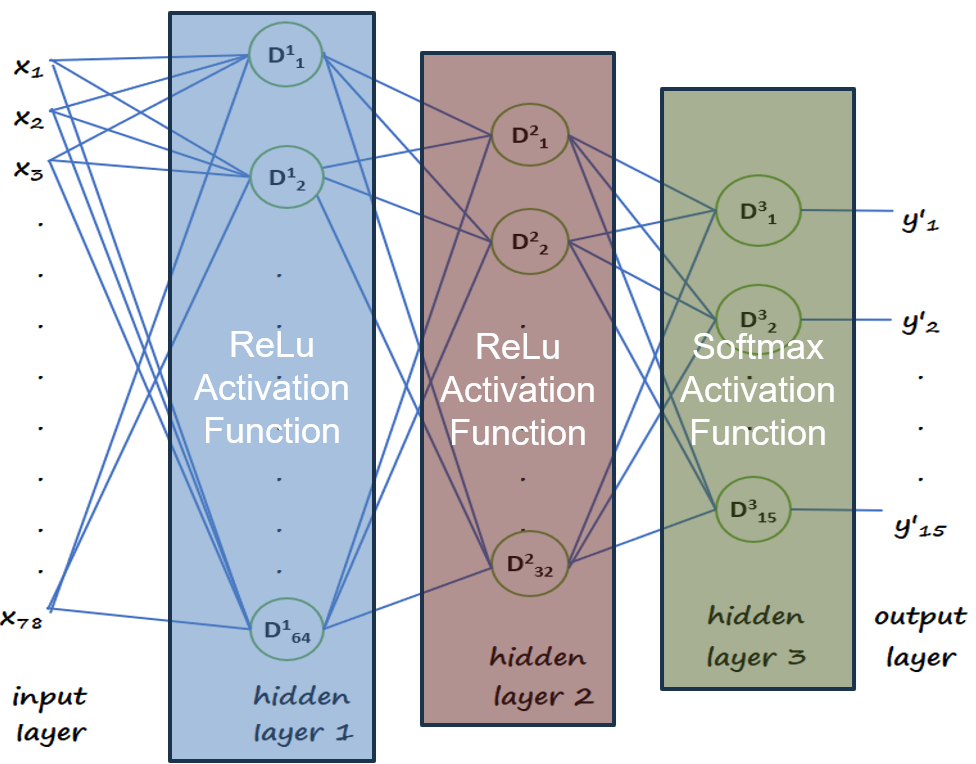}
    \caption{2-Layer FC with Softmax Layer DNN for Attack Classification}
    \label{fig:DNN}
\end{figure}

Based on the CICIDS-2017 dataset literature by \cite{sharafaldin2018CICIDS-2017}, it has extracted features required for machine learning algorithms. Taking advantage of it, this paper focuses on data augmentation and classification. Two model designs are proposed in this paper. IDS1 with binary classification as the base unit and IDS2 with double balancing and categorization as the base unit. Each of them has three parts. data pre-processing, base unit, and classification unit. The DNN in the base and classification unit has two fully-connected layers and a drop-out layer shown in Figure \ref{fig:DNN}. It is followed by the Sigmoid and Softmax layer for IDS1 and IDS2 models respectively. The fully-connected layer 1 has 64 neurons and the rectified linear unit (ReLu) activation function. The fully-connected layer 2 has 32 neurons and ReLu activation function.

\subsection{Data Pre-processing: Balancing and Standardization}
\begin{figure}[H]
    \centering
    \includegraphics[width=8cm,height=7cm]{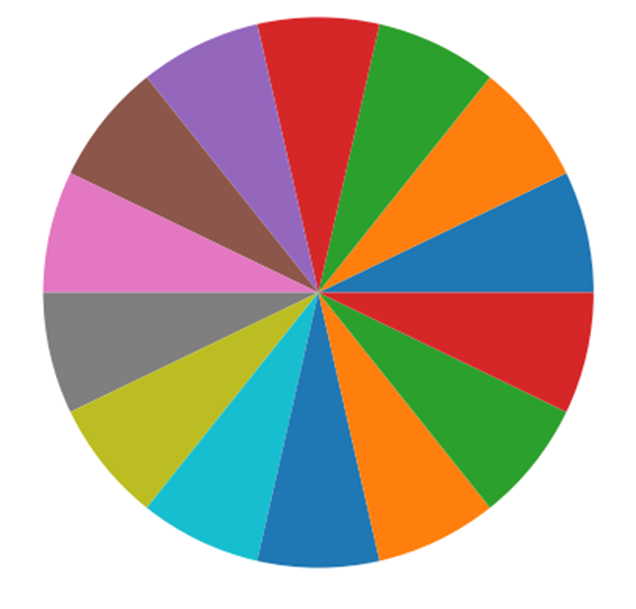}
    \caption{CICIDS-2017 After Balancing}
    \label{fig:BalancedDS}
\end{figure}

The data pre-processing step is common to both the proposed models. It handles the missing data by identifying the flows with NaN values and the values tending to negative and positive infinity, and removing them.
The class distribution is analyzed in Figure \ref{fig:attack-distri} and \ref{fig:initialDS}. It shows that CICIDS-2017 is highly imbalanced, with 80.32\% of Benign flows and three minority classes with much below 0.0001\% of total flows. In the previous phase of this project, it was evident that the little boost of minority classes by oversampling and under-sampling of benign did not fetch good F1-Scores required for IDS. The high accuracy is skewed toward abundant Benign class. It motivated the enhancement and experiment with data balancing in IDS1 and IDS2. SMOTE utilizes the nearest-neighbor algorithm to over-sample minority classes (like Heartbleed and Infiltration) and under-sample abundant classes like Benign and DDoS. The optimum level is picked for this purpose. SMOTE brings up the minority classes and downsizes the abundant classes to the optimum level. Thereby reaching a balance between all classes as shown in Figure \ref{fig:BalancedDS}. 
The balanced data is standardized using Sci-Kit Learn Preprocessing library StandardScaler. It scales the feature values between $-1$ to $1$ and the feature mean of zero. 
The nominal labels are converted to numerical labels. One-hot encoding is performed on the numerical labels using Keras Utility \textit{to\_categorical}. This is required for the final classification layer (Sigmoid/Softmax) in DNN.

\begin{figure*}[htbp]
    \centering
    \includegraphics[width=16cm,height=8cm]{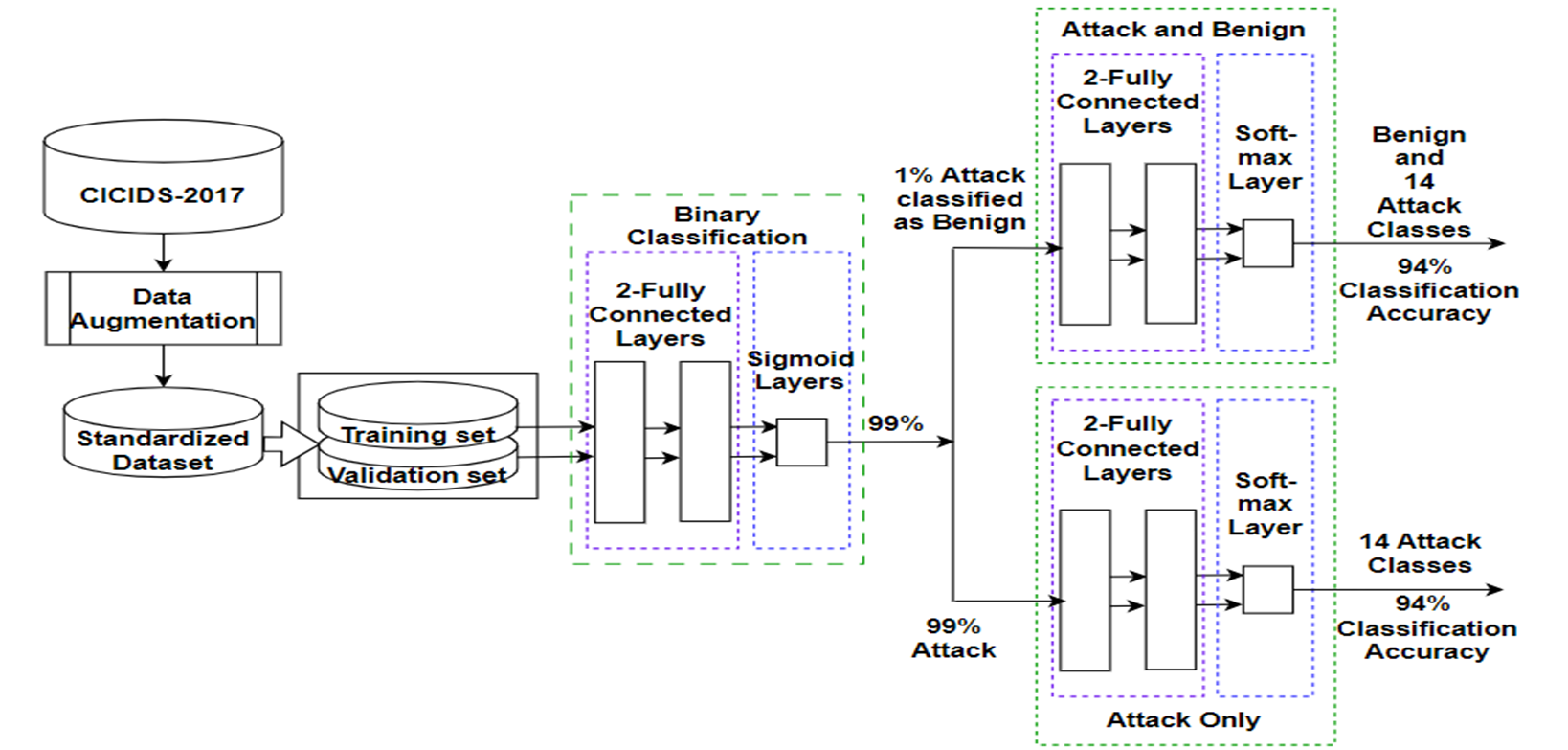}
    \caption{IDS1: Proposed Design with Binary Classification}
    \label{fig:IDS1}
\end{figure*}

\section{IDS1}
The design is depicted in Figure \ref{fig:IDS1}. It is divided into three parts. It is focused on data balancing, binary classification of benign and attack flows, and attack flows classification. 

\subsubsection{Data Balancing}
The examples are declassified in this unit into two classes: Benign and Attack. The optimum level for data balancing is 600,000 for the Binary Unit, and 250,000 for the Attack units. Balancing is achieved by under-sampling Benign class examples and over-sampling the attack classes to the optimum level. The resulting dataset is split into train and test sets in a 4:1 ratio. 

\subsubsection{Binary Unit}
The Binary Unit (BU) takes the input of train and test data and is trained. This unit has achieved 99\% accuracy and F1-scores. This means that 1\% of benign flows are classified as attacks termed as False Negatives (FN) and 1\% of attack flows classified as Benign are termed as False Positives (FP).  

\subsubsection{Classification}
False Positives are not acceptable in IDS. To deal with the high FP number, IDS1 has two classification units. One is trained with Benign and 14 attack classes referred to as the BAC unit, and the other with only attack classification referred to as the AC unit. 
In the BAC unit, the training and validation accuracy and F1-score of 94\% is achieved. It is expected to identify 0.94\% of residual attacks and enhance the attack identification rate. As a side-effect, it will add 0.94\% of FN in IDS1. 
The AC unit is trained to classify 14 attack classes with 94\% accuracy and F1-Scores. 
The final Attack identification accuracy of 99.94\% and classification accuracy of 94\% is achieved in IDS1 design.

\section{IDS2}

In IDS2 design, the categorization approach is explored as proposed by \cite{wang2023res-TransBiLSTM_imb}. The categorization is conducted by putting similar feature minority classes into one category. DDoS and PortScan are among the abundant classes assigned to the individual category. Even though DoS-Hulk is an abundant class, corresponds to DoS-type attacks having similar features as other minority DoS classes (Slowloris, Slowhttptest, and GoldenEye). Thus, all four are categorized in a single DoS category. The details are shown in Figure \ref{fig:Double}

\begin{figure}[htbp]
    \centering
    \includegraphics[width=8.5cm,height=8cm]{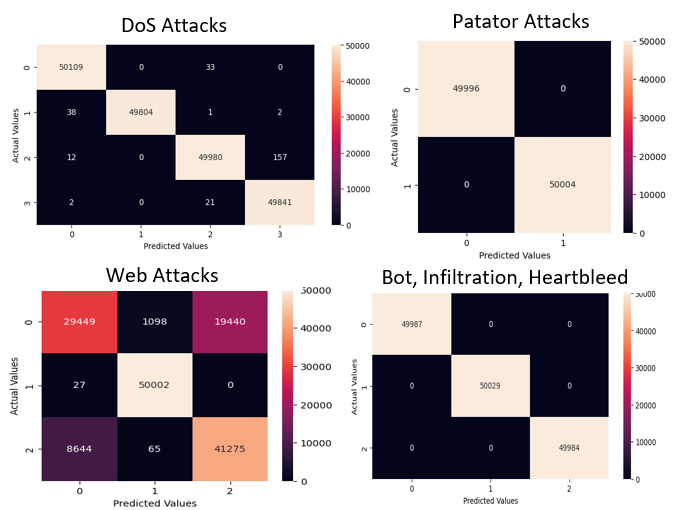}
    \caption{Sub-categorical Attack Confusion Matrix }
    \label{fig:sub-cat-ConfMat}
\end{figure}

\subsubsection{Double-Balancing of Data}

\begin{figure*}[htbp]
    \centering
    \includegraphics[width=14cm]{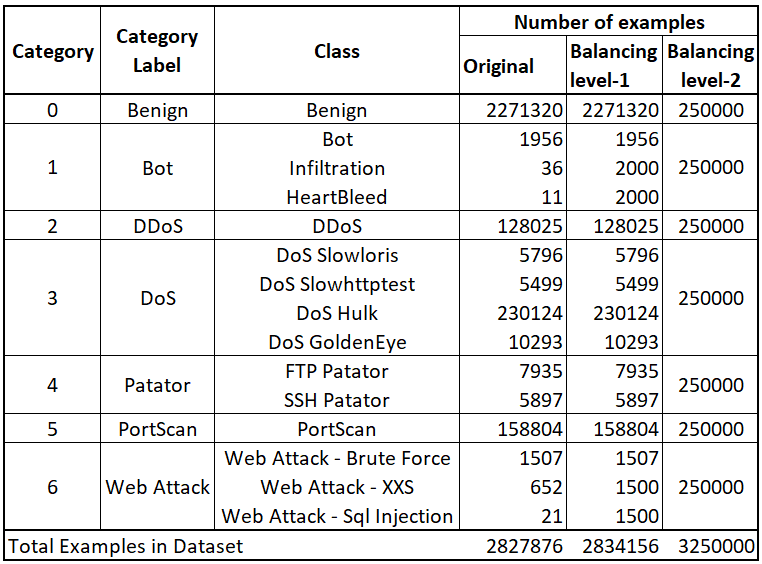}
    \caption{Double Balancing in Categorization}
    \label{fig:Double}
\end{figure*}

In IDS1, it is understood that the equal balancing of all the classes in the dataset plays a vital role in training and validation accuracy. Thus, the minority sub-classes within the attack category are balanced before categorization termed level-1 balancing. The second level of balancing is done after the categorization. The number of examples in the original dataset, level-1 and level-2 balancing are shown in Figure \ref{fig:Double}.

Level-1 Balancing within DoS and Patator Categories is not done to assess accuracy comparison between single- and double-level balancing. 

\subsubsection{Categorization Unit}
The dataset is divided into seven broad categories: Benign, Bot, DDoS, DoS, Patator, PortScan, and Web Attack.  After level-2 Balancing, the categorized dataset is splitted into training and testing sets. It is input to the Categorization Unit for Categorization. The categorization is performed at 99\% accuracy and 99\% F1-Scores. It generated three output classifications, Benign, DDoS, and PortScan. The other four outputs are the categories DoS, Bot, Web Attack, and Patator, and are fed to four respective DNN units for further classification units. 

Similar experiments are performed with Train-Validation-Test Split. The accuracy achieved was 98\%. It has F1-scores of 94\% and 93\% for the Patator and Web Attacks categories. 

\subsubsection{Classification Unit}
This unit has 4 DNN units, DNN-1 for DoS Classification, DNN-2 for Bot Classification, DNN-3 for Web Attack Classification, and DNN-4 for Patator Classification. Each classified the sub-classes with almost 100\% accuracy except Web Attack, which has 80\% accuracy. The confusion matrix for the four units is shown in Figure \ref{fig:sub-cat-ConfMat}.

\begin{figure}[htbp]
    \centering
    \includegraphics[width=8.5cm,height=8cm]{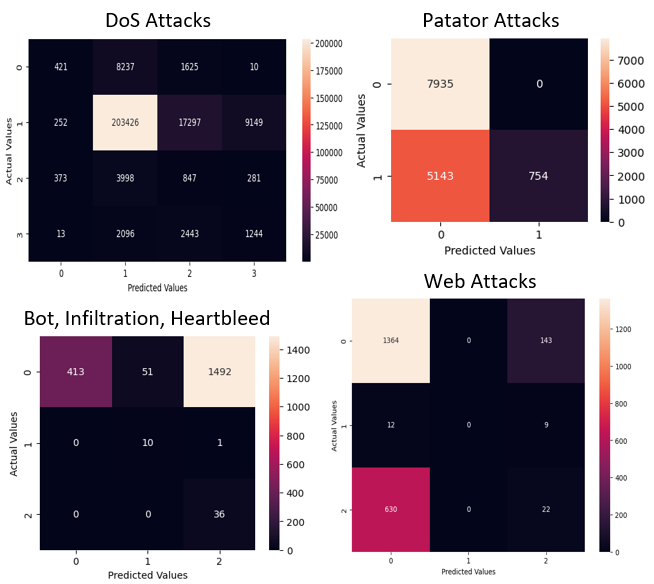}
    \caption{Sub-categorical Attack Confusion Matrix on Original Dataset}
    \label{fig:sub-cat-ConfMatOrg}
\end{figure}

Similar experiments are performed with the Train-Validation-Test Split. The accuracy and F1-Scores achieved for DoS, Bot, and Patator was almost 100\% and Web Attack was 79\%. It has F1-scores of 94\% and 93\% for the Patator and Web Attacks categories. And, with the original dataset without balancing for DoS, Bot, Web Attack and Patator showed accuracies of 83\%, 23\%, 4\%, and 50\%. The heat maps of the confusion matrix are shown in Figure \ref{fig:sub-cat-ConfMatOrg}. This signifies the over-fitting caused by balancing data. 

\begin{figure*}[htbp]
    \centering
    \includegraphics[width=12cm,height=3.75cm]{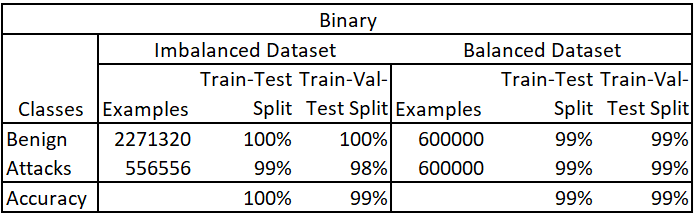}
    \caption{Effect of Balancing on Binary Classification F1-Scores}
    \label{fig:BinaryF1ScoresBalVSImb}
\end{figure*}
\subsection{Training and Testing}
Both IDS1 and IDS2 have multiple units in DNN architecture. Each unit is trained and tested individually using the same dataset with different over-sample and under-sample numbers. The designs' accuracy is calculated mathematically. 

\section{Results}
IDS1, train and test split of the balanced dataset is believed to have 
\begin{itemize}
    \item The overall Attack-Vs-Benign classification accuracy of 99.94\%\\
    (BU-Accuracy) $+$ (BAU-Accuracy*0.01)
    \item The overall Attack sub-classes classification accuracy of 94.48\%\\
    (BU-Accuracy*AU-Accuracy) $+$ (BU-Accuracy*BAU-Accuracy*0.01)
\end{itemize} 

\begin{figure}[H]
    \centering
    \includegraphics[width=8.5cm,height=10cm]{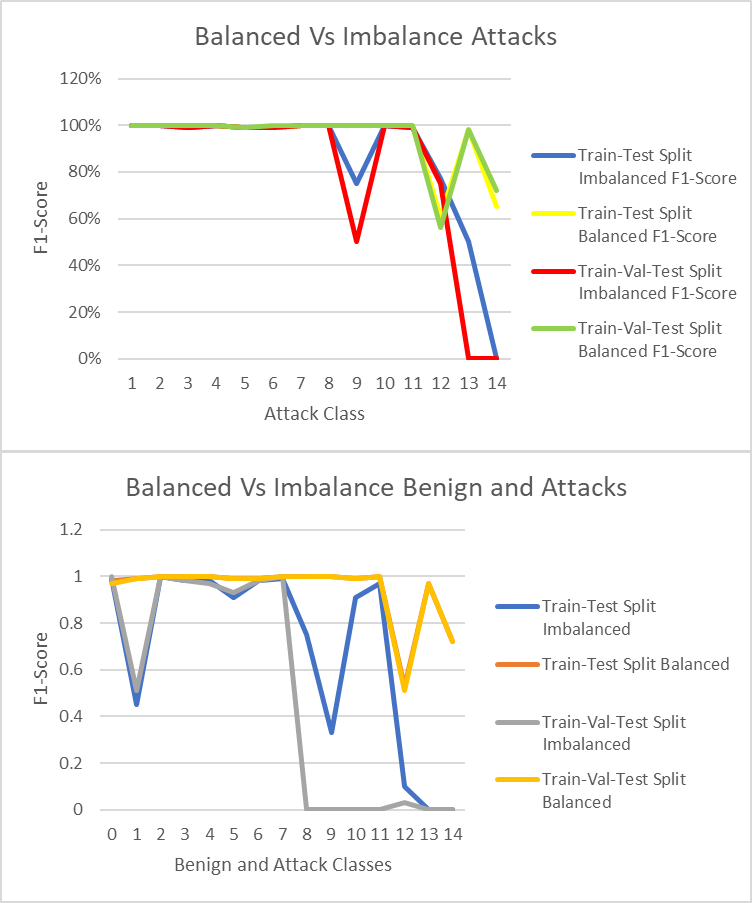}
    \caption{Balanced Vs Imbalanced F1-Scores for Attack Unit (AU) and Benign and Attack Unit (BAU)}
    \label{fig:BalVsImb}
\end{figure}

Based on IDS1 and IDS2, a few general experiments are performed to compare the effect of balancing on the Binary Unit training of the same architecture shown in Figure \ref{fig:BinaryF1ScoresBalVSImb}, and attacks classifications unit shown in Figure \ref{fig:BalVsImb}. While the Binary unit does not show the balancing benefit, the attack sub-classes classification units show improvement in F1-Scores.

\section{Conclusion}
The various experiments conducted show that data augmentations are required in imbalanced datasets. At the same time, it cannot be conducted as a thumb of rule. In the Binary Unit, data augmentation may not be required. The testing on the original dataset makes it evident that data augmentation is causing overfitting. The fact that the Data Augmentation technique SMOTE is oversampling using the nearest neighbor algorithm based on Euclidean distance to generate more examples. This type of data augmentation is not appreciable. In Res-TranBiLSTM, SMOTE-ENN is used which deletes the overlapping examples. But still causes overfitting. 

In State-of-the-Arts designs like Res-TranBiLSTM by \cite{wang2023res-TransBiLSTM_imb}, RTIDS by \cite{wu2022rtids}, IDS\_INT by \cite{ullah2023ids} and CNN-transformer hybrid by \cite{yao2023cnn}, a common architecture used is transformer considering the actual network. But all are tested on multiple datasets including CICIDS-2017. Interestingly, CICIDS-2017 has stripped off timestamp information which corresponds to temporal feature. It is expected to be present in the actual network environment, where these models should perform better.

\section{Future Work}
The data augmentation using SMOTE is counter-productive. A larger dataset like CICIDS-2018, or a mix of IDS datasets, is one alternative to be experimented with. The ADASYN, Borderline-SMOTE, and SVM-SMOTE are other alternatives to be experimented with. 
Support Vector Machine algorithm could be used to identify the Benign class using one-versus-all method.  
The neural networks have strength in extracting spatial and temporal features. Using CNN- and transformer-based architectures, it can be explored. 
The current architecture is incapable of identifying Zero-day and unknown attacks. Transformer, Bidirectional-LSTM, and CNN-based models are suggested based on the related work to achieve detection capabilities in real-world scenarios. Further, based on related work and these experiments, it is concluded that the Intelligent IDS is required to be a logical design of the hybrid model using machine learning algorithms and neural networks to make it capable of identifying benign, known, unknown, and zero-day attacks.

\section{Acknowledgments}
The author of this paper like to acknowledge the contributions of Dr. Jugal Kalita who has provided deep knowledge of multiple neural network architectures, their histories and evolution, multiple remarkable papers, detailed feedback on this work, encouragement, support on various topics, and challenges faced during the project execution. 

\bibliography{ref}

\end{document}